\documentclass[aps,twocolumn,showpacs,floatfix]{revtex4}

\usepackage[xdvi]{graphicx}
\usepackage{pdfpages}
\usepackage{latexsym}
\usepackage{amsbsy}\usepackage{amsmath}
\usepackage{amssymb}
\usepackage{lineno}
\usepackage{multirow}

\begin{document}

\date{\today}

\title{Investigation of the spin dynamics of quantum spin dimers with
  Dzyaloshinsky-Moriya interaction}  

\author{R. Wieser$^{1,2}$, R. S\'anchez Gal\'an$^3$}
\affiliation{1. School of Physics and Optoelectronic Engineering,
 Nanjing University of Information Science and Technology, Nanjing
 210044, China \\
2. Jiangsu Key Laboratory for Optoelectronic Detection of Atmosphere and Ocean,
Nanjing University of Information Science and Technology, Nanjing
210044, China \\
3. 4i Intelligent Insights, Tecnoincubadora Marie Curie, PCT Cartuja,
41092 Sevilla, Spain}

\begin{abstract}
We investigate the ground state configuration and spin dynamics of a
quantum spin spiral dimer when a magnetic field is applied to one of
its constituent spins. By adiabatically changing the magnetic field, it is
possible to change the non-magnetic ground state to a classical spiral
state, which oscillates periodically to its inverted classical spiral
state configuration. This oscillation can be halted at any time,
leading to the possibility of manipulating the quantum state and the
magnetic configuration of the spin dimer. Notably, this idea is not limited 
to spin dimers and can also be extended to quantum spin chains with an even 
number of spins.   
\end{abstract}

\maketitle

Spin dimers are pairs of magnetic atoms, ions or more generally, adjacent spin sites that interact via magnetic forces with each other. They form spin dimer compounds with unique properties that distinguish them from non-dimerized magnetic lattices: the whole phase diagram can be explored with the tuning of an external magnetic field, anisotropic effective exchanges can be engineered between spin dimers (Heisenberg, Ising or XY-like) and the interdimer exchange energy suppresses the effects of symmetric anisotropies such as the dipolar coupling \cite{samulon2008ordered}. 

Moreover, spin dimers provide a simple model for studying more complex magnetic phenomena, such as quantum entanglement, quantum phase transitions and low-dimensional magnetism.
These magnetic interactions play a significant role in data storage, spintronic devices, quantum computing, and quantum information. The trend is toward smaller structures and objects, such as nanoparticles  
\cite{hinzkePRB98,nowakPRB05,usadelPRB06,sukhovJMMM08}, molecules
\cite{boganiNatMat08,gaitaarino19,slotaAMR20,morenopinedaNatRevPhys21},
or atomic clusters
\cite{khajetooriansSCIENCE11,khajetooriansNATPHYS12,yangPRL17,
  yangNATCOMM21}. While larger structures, such as magnetic films or layer systems can be well described with classical physics
\cite{wieserPRB04,wieserPRB06,wieserPRL08,wieserPRB09,wieserPRB10,wieserPRL11,
  wieserEPL12}, this changes for small magnetic systems, such as
magnetic molecules or small clusters, when well decoupled from
their environment
\cite{wieserNJP13,wieserEPJB15,wieserJPComm19}. Here, the classical description fails, and we have to replace it with a quantum mechanical
portrayal. This fact opens the possibility of using such magnetic
systems for quantum computing or quantum information processes.

Spin $S=1/2$ dimer systems are predominantly found in Cu$^{2+}$-based compounds since the Cu$^{2+}$ ions with an electronic configuration of $3d^9$ can readily pair up to form a spin dimer \cite{deisenhofer2006structural,jaime2004magnetic,janson2011cacu,LAN2022123039,sasago1997temperature,sankar2014crystal,takagi2006crystal,urushihara2020crystal,okada2016quasi}.

These spin dimer systems have offered many fascinating examples in condensed matter physics. For instance, Bose-Einstein condensation of magnons was found in \cite{jaime2004magnetic}  and  \cite{okada2016quasi}, and Wigner crystallization of magnons, along with distinct magnetization plateaus, can occur in SrCu$_2$(BO$_3$)$_2$ \cite{kageyama1999exact}.

\begin{figure}[ht]
  \begin{center}
    \includegraphics*[width=7.0cm,bb = 150 480 550 770]{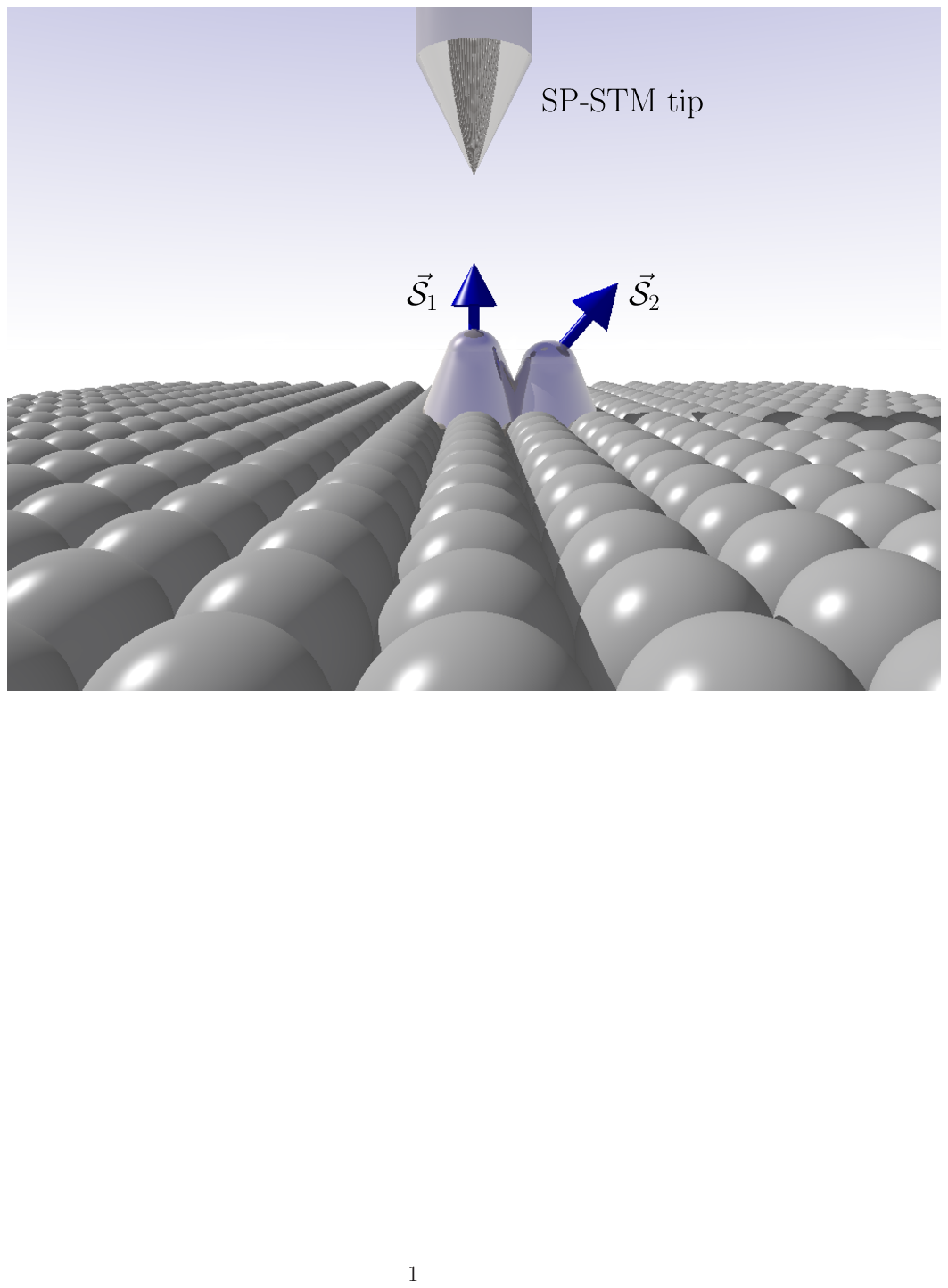}
  \end{center}
  \caption{(color online) A possible experimental setup for the spin
    dynamics described in the manuscript. The Spin-Polarized Scanning
    Tunnelling Microscope (SP-STM) tip is used to manipulate the spin
    $\vec{S}_1$. The spin dimer itself shows a non-collinear spin
    structure caused by the exchange and Dzyaloshinsky-Moriya
    interactions.} 
  \label{f:pic1}
  \end{figure}

This publication focuses on the description of the periodic
longitudinal spin dynamics of spin-spiral spin chains and dimers, as well as on the
possibility of manipulating and using these dynamics to generate or
fixate certain quantum states. The first part of the study focuses on spin dimers, providing a detailed theoretical framework. The second part extends the analysis to longer spin chains with an even number of spins, demonstrating the feasibility of applying the same principles. 

The ability to leverage the spin dynamics of spin-spiral dimers and chains is grounded in the following key ideas:

\begin{enumerate}
\item The quantum mechanical spiral state differs from the classical
spiral state, i.e., the classical spin spiral
\cite{voelkelAnnPhys93,vedmedenkoPRB07,soltaniISRNCMP11,hauptmannNatComm20}.
This is analogous to the quantum mechanical antiferromagnetic state,
which differs significantly from the classical N{\'e}el state. The
even-odd effect, which can be observed in antiferromagnetic clusters
\cite{machensPRB13}, is also found in spin spiral clusters.  
\item In quantum mechanical spin systems, there are additional
dynamics or effects besides the classical ones: precession and
transverse relaxation. For example, quantum tunneling
\cite{wernsdorferSCIENCE99,nishinoPRB01,pokrovskyPRB03,pokrovskyPRB04}
, longitudinal relaxation 
\cite{galkinaPRB13}, and entanglement between different spins
\cite{wieserEPJB15,wieserJPComm19,molmerPRL99} can be found in quantum
mechanical spin systems, but not in classical ones. In particular,
entanglements play a significant role in quantum computers
\cite{sorensenPRL99,lanyonScience11,montanaroNPJ16,albashRMP18,mcardleRMP20,
blaisRMP21} or quantum teleportation \cite{braunsteinRMP05,
gisinNaturePhotonics07}. Longitudinal relaxation, on the other hand, plays a
major role in quantum state transfer  
\cite{joelAJP13,marchukovNATURECOMM16,lorenzoIJQI17,bazhanovSR18,
  shanSCIREP18,wangNATUREPhys19}.     
\item The ability to artificially generate, study, and manipulate
classical or quantum mechanical spin clusters or chains. It does not
matter whether the resulting cluster is collinear or non-collinear,
i.e., with a spiral. Most of these experiments are carried out using spin-polarized scanning tunnelling microscopes
\cite{meierSCIENCE06,serrateNATNANO10,khajetooriansSCIENCE11,
  khajetooriansNATPHYS12,yangPRL17,yangNATCOMM21,zhangNatureChem22}
or the tip of a Magnetic Exchange Force Microscope (MEx-FM)
\cite{kaiserNATURE07,wieserNJP13}.
\item Recent experiments show that artificially placed atoms or
clusters can be used as Qubits and can be controlled by a combination
of Electron Spin Resonance (ESR) and Spin-Polarized Scanning Tunneling
Microscope (SP-STM)
\cite{wangNPJQI23,zhangNatureChem22,hanzeSCIADV21,yangNATCOMM21}. This publication shows an alternative way to use such atomic clusters for quantum computing.   
\end{enumerate}

More specifically, we consider a spin dimer described by a Heisenberg model with
exchange and Dzyaloshinsky-Moriya interactions and a local coupling to
an external magnetic field. In other words, the Hamiltonian describing
the model is    
\begin{eqnarray} \label{HamDMI}
  {\cal H} = -J\vec{S}_1\cdot\vec{\cal S}_2 + D\vec{x}\cdot\left(
  \vec{\cal S}_1\times\vec{\cal S}_2 \right) - B_1 {\cal S}_1^z \;.
\end{eqnarray}
The spins are $S = 1/2$ quantum spins. The exchange coupling between
the two spins is assumed to be ferromagnetic $J > 0$. In the
following, $J$ is used as a reference for the other interactions,
meaning we consider $D$ and $B_1$ in units of $J$. The
Dzyaloshinsky-Moriya interaction creates, together with the exchange interaction, a spin spiral with spiral angle $\theta =
\mathrm{arctan}(D/J)$. In this particular case, the
Dzyaloshinsky-Moriya vector $\vec{D} = D\vec{x}$, is oriented in
the $x$-direction. Therefore, the spin spiral is in the $yz$-plane, and
the rotation occurs clockwise if $D > 0$ and counter-clockwise if $D <
0$. The last term, $B_1 = \mu_S B$, describes the effect of an
external magnetic field or spin-polarized current on the first
spin. The second spin is not affected by this external disturbance. 

Using exact diagonalization, one can find the ground state configuration corresponding to the Hamilton operator
${\cal H}$. In particular, when $B_1 = 0$, the ground
state denoted $|{\cal GS}\rangle$ is given by 
\begin{eqnarray}
  |{\cal GS}\rangle =
  \frac{1}{\sqrt{2}}\!\Big(|\Psi_{S,\uparrow}\rangle +
  |\Psi_{S,\downarrow}\rangle \Big) \;,
\end{eqnarray}
with $|\Psi_{S,\uparrow}\rangle$ and $|\Psi_{S,\downarrow}\rangle$
being the two possible classical spin spiral states corresponding to
the Hamilton operator ${\cal H}$. If the external magnetic field is zero, the first spin could be oriented in any direction within the $yz$-plane. The second spin is then twisted
clockwise or counter-clockwise with respect to the first spin by the spiral
angle $\theta$. With a non-vanishing external magnetic field $B_1 \neq
0$, the first spin tends to orient towards the magnetic
field. Therefore, and without loss of generality, the first spin in
$|\Psi_{S,\uparrow}\rangle$  shall be oriented in the
$+z$-direction. In the second spiral state
$|\Psi_{S,\downarrow}\rangle$, the first spin is oriented in the
$-z$-direction. Actually, the classical spiral states can be written as product states of two coherent spin states:  
\begin{subequations}
\begin{eqnarray}
|\Psi_{S,\uparrow}\rangle &=& |\uparrow\rangle \otimes
\left(\cos\frac{\theta}{2}|\!\uparrow\rangle +
\sin\frac{\theta}{2}|\!\downarrow\rangle \right) \;, \\
|\Psi_{S,\downarrow}\rangle &=& |\downarrow\rangle \otimes
\left(\cos\frac{\theta}{2}|\!\downarrow\rangle +
\sin\frac{\theta}{2}|\!\uparrow\rangle \right) \;.  
\end{eqnarray}
\end{subequations}
Within these states, $\theta$ is the angle of the spin spiral. As
mentioned above, these states describe classical spin spirals. We must
mention that $|\Psi_{S,\uparrow}\rangle$ and
$|\Psi_{S,\downarrow}\rangle$ are not the ground state
configuration. The ground state is a superposition of these two states. In fact, the ground state is
entirely entangled and shows no magnetic order i.e., if $\vec{\cal S}_{1,2} $ denotes the total magnetization operator of the spin dimer, then
\begin{eqnarray}
  \langle {\cal GS}| \vec{\cal S}_{1,2} |{\cal GS} \rangle =  0.
\end{eqnarray}
\begin{figure*}[!htbp]
  \begin{center}
    \includegraphics*[width=7cm,bb = 75 460 510 765]{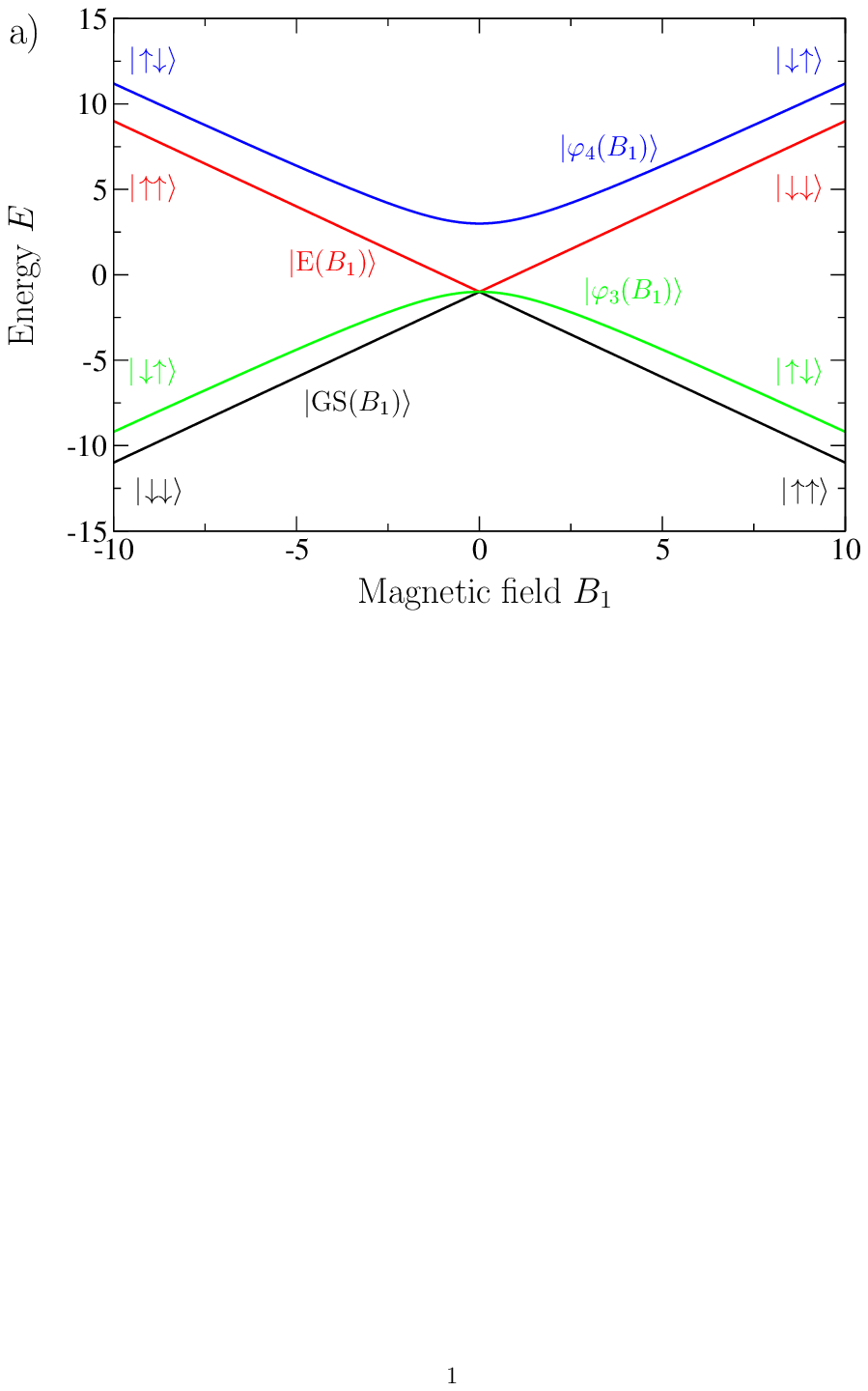}
    \includegraphics*[width=7.4cm,bb = 75 475 510 765]{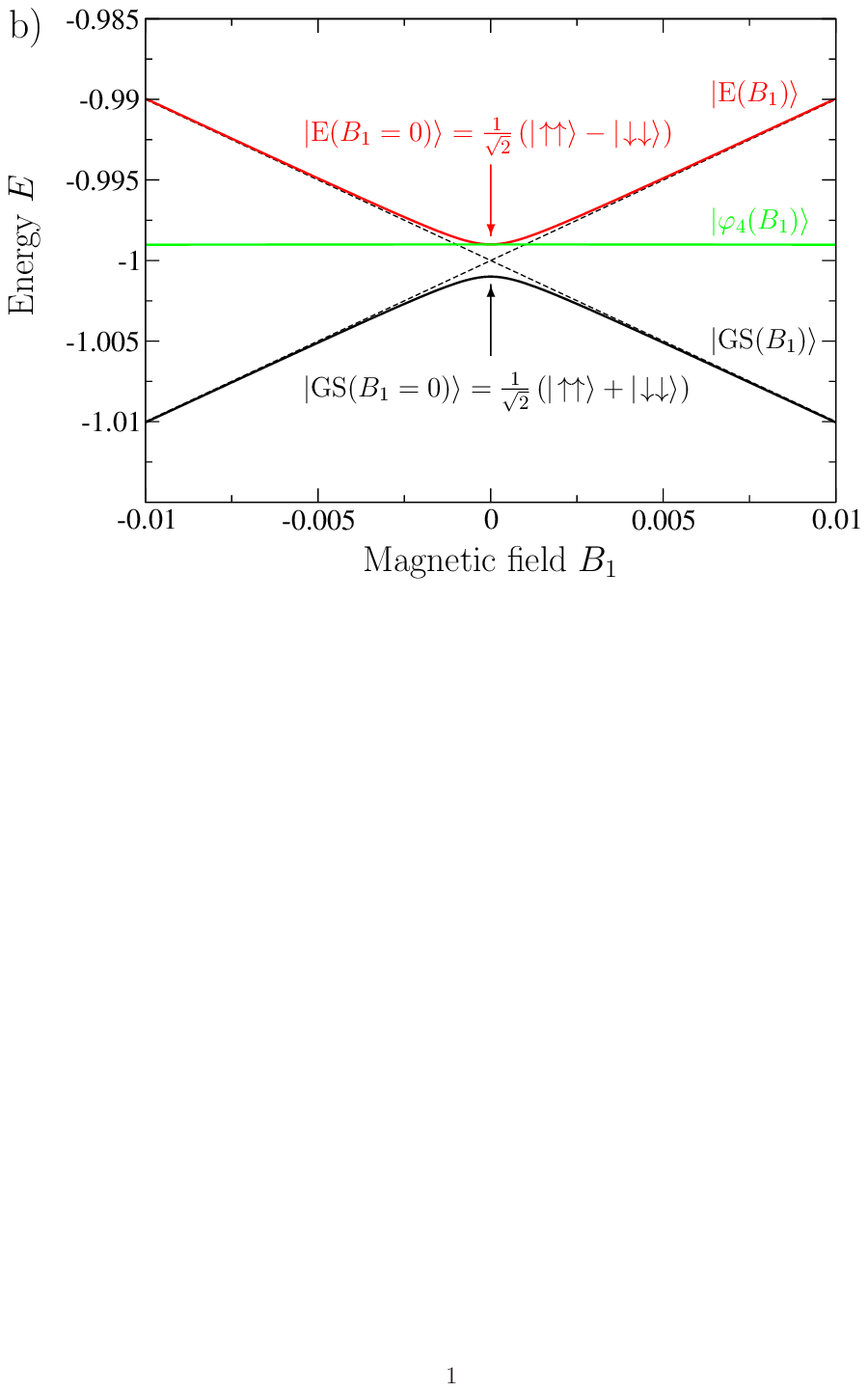}
  \end{center}
  \caption{(color online) The energies of the eigenstates
    $|\varphi_n(B_1)\rangle $, $n \in \{1,2,3,4\}$ as a function of the
    magnetic field $B_1$. In the limits $B_1 \rightarrow \pm \infty$, the
    eigenstates $|\varphi_n\rangle$ become the classical ferromagnetic
    $|\!\uparrow\uparrow\rangle$, respectively,
    $|\!\downarrow\downarrow\rangle$ and
    antiferromagnetic $|\!\uparrow\downarrow\rangle$, respectively,
    $|\!\downarrow\uparrow\rangle$ spin spiral states. 
    The first eigenstate $|\varphi_1\rangle = |\mathrm{GS}\rangle$ and
    the second eigenstate $|\varphi_2\rangle = |\mathrm{E}\rangle$, are the most relevant for the spin dynamics under
    consideration. a) general overview and b) detailed view of the
    bandgap at $B_1 \approx 0$.} 
  \label{f:pic2}
\end{figure*}

\begin{figure}[!htbp]
  \begin{center}
    \includegraphics*[width=7cm,bb = 75 460 510 765]{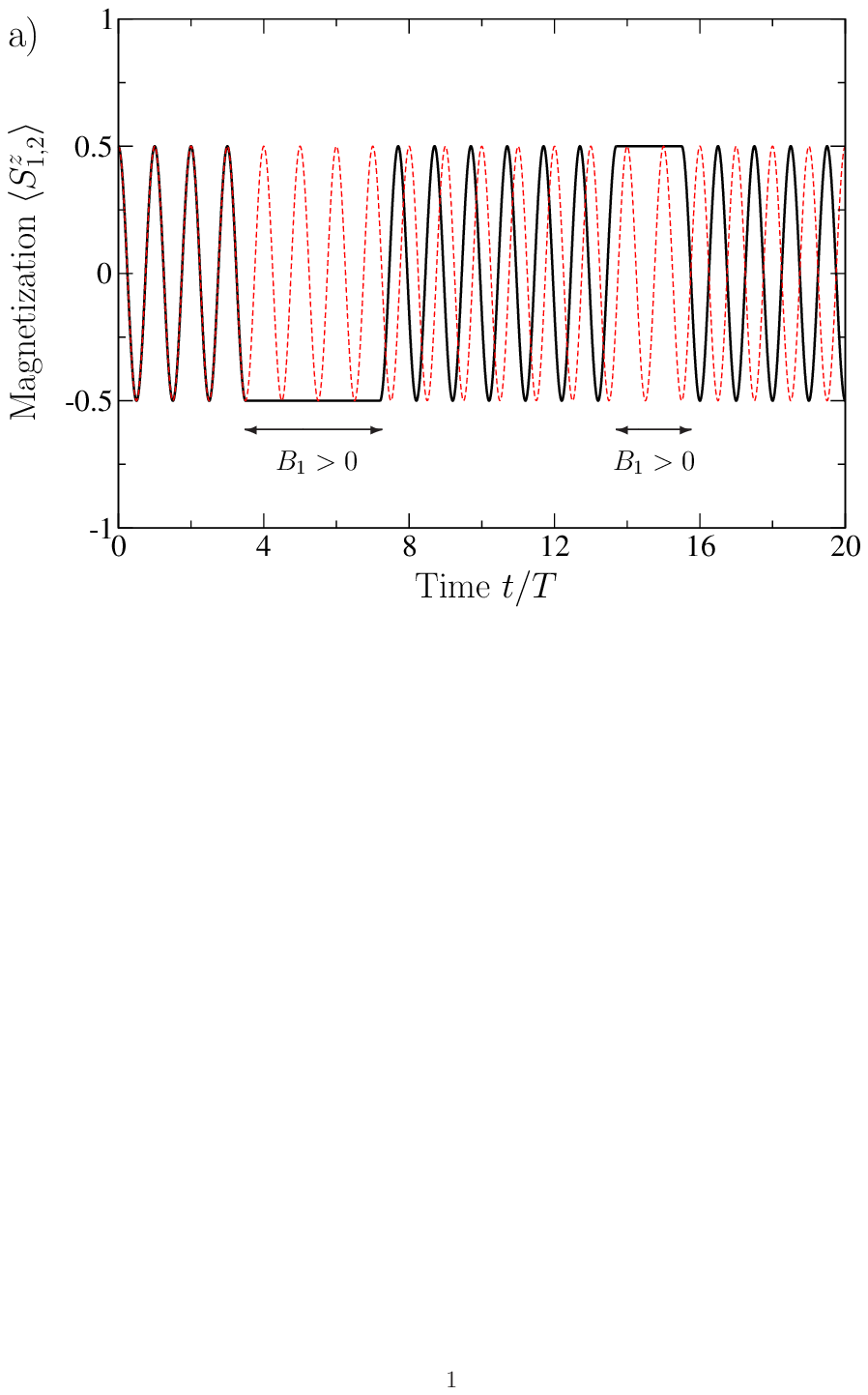}
    \includegraphics*[width=7cm,bb = 75 460 510 765]{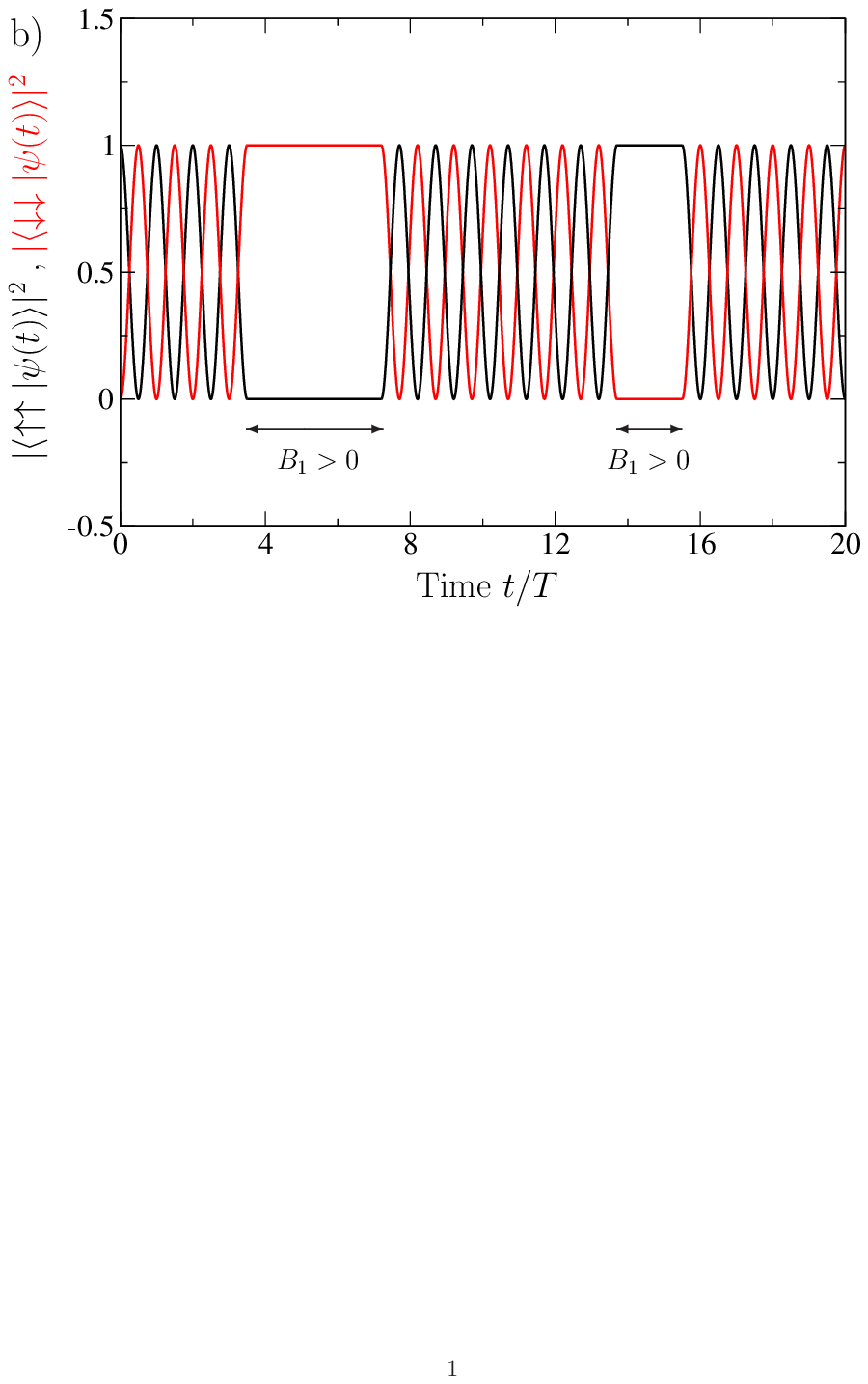}
  \end{center}
  \caption{(color online) The first of two considered scenarios: the
    oscillation of $|\psi(t)\rangle$ between the two classical spiral
    states stops when $|\psi(t)\rangle$ is in one of these classical
    states. Figure a) shows the magnetization $\langle S_{1,2}^z
    \rangle$ and figure b) the probabilities to find $|\psi(t)\rangle$ in one of the two classical spiral states as a function of time,
    with $T = \pi \hbar/ J_x$ being the time of one complete oscillation. The
    orientation of the external magnetic field $B_1$ is not essential, it can be in the $\pm z$-direction. The two arrows mark the times when the external magnetic field is switched on $B_1 > 0$. Outside these
    time intervals, the external magnetic field is switched off, $B_1
    = 0$, allowing the oscillation of $|\psi(t)\rangle$ between the two classical spiral states. The release time is arbitrary and allows
    any possible phase shift with respect to the oscillation of the
    unperturbed spin dimer (red dashed line).}     
  \label{f:pic3} 
\end{figure}
Due to the non-collinear structure of the spin spirals, calculations
involving the Hamilton operator ${\cal H}$ are relatively
complicated. To simplify the calculations, it makes sense to transform the coordinate system to local coordinates
\cite{voelkelAnnPhys93,wieserJPCM19}. After this coordinate
transformation, each spin is oriented along its $z$-axis within the
new coordinates and the structure is collinear. The Hamilton operator in these coordinates is given by:
\begin{eqnarray}
  \mathrm{H} = -\tilde{J} \vec{S}_1 \cdot \vec{S}_2 + J_x
  S_1^xS_2^x - B_1S_1^z \;,
\end{eqnarray}
with $\tilde{J} = J/\cos[\mathrm{arctan}(D/J)]$, and
$J_x = \tilde{J} - J$. To achieve this transformation, we had to rotate the spins, with the rotation operators defined in the supplemental material. 

Due to the transformation, the Heisenberg model with the Dzyaloshinsky-Moriya interaction becomes an XXZ model. The XXZ model, along with the more general XYZ model, and their associated dynamics have been extensively studied. Several notable publications explore the spin dynamics within the XXZ model framework. For instance, Marchukov et al. proposed a coherent spin transistor based on the XXZ model that operates via longitudinal relaxation \cite{marchukovNATURECOMM16}. More broadly, the XXZ and XYZ models are pivotal in studies on quantum state transfer \cite{joelAJP13,marchukovNATURECOMM16,lorenzoIJQI17,bazhanovSR18, shanSCIREP18,wangNATUREPhys19}. The XXZ model has also been employed to describe quantum domain walls \cite{gochevPL81,gochevPL84} and their interactions with spin waves \cite{yuanJPSJ06}. Furthermore, when the Dzyaloshinsky-Moriya interaction is incorporated along the $z$-direction, the XXZ model can be used to implement a quantum SWAP gate \cite{gurkanPhD2010}.

The ground state $|\mathrm{GS}\rangle$ of the investigated dimer is the superposition of the two ferromagnetic states, which represent
the two classical spiral states, $|\Psi_{S,\uparrow}\rangle$ and
$|\Psi_{S,\downarrow}\rangle$ in local coordinates, 
\begin{eqnarray} \label{Grundzustand}
  |\mathrm{GS}\rangle =
  \frac{1}{\sqrt{2}}\left(|\!\uparrow\uparrow\rangle +
  |\!\downarrow\downarrow\rangle \right) \;.
\end{eqnarray}
The energy of the ground state in this new coordinate system is
\begin{eqnarray}
\langle\mathrm{GS}| \mathrm{H} |\mathrm{GS} \rangle =
-\tilde{J} - J_x \;.
\end{eqnarray}

Exact diagonalization provides the four eigenstates and eigenenergies
of the Hamilton operator $\mathrm{H}$. However, next to the ground state
$|\mathrm{GS}\rangle$, only the excited state 
\begin{eqnarray} \label{Anregung}
  |\mathrm{E} \rangle = \frac{1}{\sqrt{2}}
    |\left(|\!\uparrow\uparrow\rangle -
    |\!\downarrow\downarrow\rangle \right) \;,
\end{eqnarray}
with eigenenergy $\langle
\mathrm{E}|\mathrm{H}|\mathrm{E}\rangle = -J + 
J_x$, is of genuine interest to our study. Since there are two $S =
1/2$ spins, we expect four different eigenenergies. In fact, these four 
eigenenergies are plotted in Fig.~\ref{f:pic2} as a function of the local
magnetic field $B_1$. Together with the ground state
$|\mathrm{GS}\rangle$, the excited state $|\mathrm{E}\rangle$ forms an
adiabatic system with an energy gap of $\Delta = 2J_x$, see
Fig.~\ref{f:pic2}b). When increasing $B_1$ in the $\pm z$-direction, the ground state becomes one of the two classical spin spirals,
$|\Psi_{S,\uparrow}\rangle$ or $|\Psi_{S,\downarrow}\rangle$. The energy
of the corresponding excited spin spiral state
$|\mathrm{E}\rangle$ increases
correspondingly. This energy configuration allows an adiabatic change
of the ground state away from the superposition
$|\mathrm{GS}\rangle$ with no magnetic order $\langle \vec{S}_n
\rangle = 0$ towards one of the two classical spin spiral configurations,
$|\!\uparrow\uparrow\rangle$ or $|\!\downarrow\downarrow\rangle$.

When releasing the classical spiral state, e.g.,
$|\Psi_{S,\uparrow}\rangle$, by adiabatically switching off the external magnetic
field $B_1$, the quantum state is no longer an eigenstate of the
Hamiltonian ${\cal H}$. Under these circumstances,
$|\Psi_{S\uparrow}\rangle$ and $|\Psi_{S,\downarrow}\rangle$ represent linear combinations of excited states with the same energy $E = -J$. Therefore, we find a
periodic change of the quantum state $|\psi(t)\rangle$ between these
two states. This is similar to the periodic change of the
configurations of the ammonia molecule, with the difference that the
ammonia molecule is in the ground state \cite{cohentannoudjiBOOK1}.

The spin dynamics in local coordinates, assuming that $B_1 = 0$ and the initial quantum state is the classical spin spiral state
$|\psi(t=0)\rangle = |\!\uparrow\uparrow\rangle$, can be
computed by solving the time-dependent Schr\"odinger equation. The spin dynamics are given by the ket
\begin{eqnarray} \label{PSItime}
|\psi(t)\rangle =
e^{\frac{i\tilde{J}t}{\hbar}}\!\left[\cos\!\left(\frac{J_xt}{\hbar}\right)
|\!\uparrow\uparrow\rangle 
- i
\sin\!\left(\frac{J_xt}{\hbar}\right)|\!\downarrow\downarrow\rangle
\right] \;. 
\end{eqnarray}
The corresponding quantum state in global coordinates is obtained by
rotating the second spin by the spiral angle $|\Psi(t)\rangle = \exp(-i\theta
S_2^y)|\psi(t)\rangle$. The general solution for a given initial state and a description of the details of the rotational transformation is given in the supplementary materials \cite{Suppl}.  

The associated magnetization dynamics are given in local coordinates
by $\langle S_{1,2}^x \rangle = \langle S_{1,2}^y
\rangle = 0$ and 
\begin{eqnarray} \label{MagTime}
\langle S_{1,2}^z \rangle = \cos\!\left(\frac{2J_xt}{\hbar}\right) \;.
\end{eqnarray}
The frequency of the oscillation, $\omega = 2J_x/\hbar$, depends on
$J_x$, and hence on the strength of the Dzyaloshinsky-Moriya interaction
$D$ concerning the isotropic exchange $J$. By changing $D$, e.g., with
an electric field \cite{siratoriJPSJ80,chenPRL15}, the frequency
$\omega$, and the resulting oscillation period $T = 2\pi/\omega$ can
be modified.    

As for the quantum state $|\psi(t)\rangle$, the corresponding result
in global coordinates can be achieved by rotating the second spin
around the $y$-axis by the spiral angle $\theta$: $\langle \vec{\cal S}_2
\rangle = R_y(\theta)\langle \vec{S}_2 \rangle$.
Equations (\ref{PSItime}) and (\ref{MagTime}) show that the quantum state
$|\psi(t)\rangle$ periodically changes between the two spin spiral
states. These spin dynamics are unusual for classical spin dynamics because
there is a longitudinal relaxation, and it does not conserve spin
length. During this oscillation, the two spins change between not
entangled and maximally entangled. The two spins are not entangled if
they are in one of the classical spin spiral states,
$|{\psi}_{S,\uparrow}\rangle$ or
$|{\psi}_{S,\downarrow}\rangle$, and maximally entangled if the
quantum state is an equal superposition between the two classical spin spiral states. The oscillation can be stopped at any time by applying the external magnetic field $B_1$ to the first spin. Even
though every possible superposition of the two classical spiral states
can be reached, two interesting scenarios can be achieved. The first of the two scenarios to be considered is to stop the oscillation at the moment when the system is in one of the classical spiral states. In this scenario, the quantum state $|\psi(t)\rangle$ stays in
the classical spin spiral state $|\Psi_{S,\uparrow}\rangle$ or
$|\Psi_{S,\downarrow}\rangle$ as long as the magnetic field $B_1$ is switched on. It is not important if the magnetic field is in the $+z$- or $-z$-direction to stop the oscillation. What is relevant is that the magnitude of $B_1$ is larger than the driving anisotropic exchange
interaction $B_1 \gg J_x$. The oscillation continues without
modification once the external magnetic field is switched off. The
only change with respect to the unaltered oscillation is a phase
difference. This phase shift can be adjusted continuously, as the
oscillation can be released at any time. The stopping time must
coincide with the condition that $|\psi(t)\rangle$ is in one of the two classical spiral states. Within this scenario, the dimer can be in one of these three states: fixed in the first classical spin spiral state 
$|\Psi_{S,\uparrow}\rangle$, fixed in the second classical spin spiral state
$|\Psi_{S,\downarrow}\rangle$ or oscillating between these two
classical spiral states. Fig.~\ref{f:pic3} shows an example of this scenario. 
\begin{figure*}[!htbp]
  \begin{center}
    \includegraphics*[width=14cm,bb = 75 460 540 765]{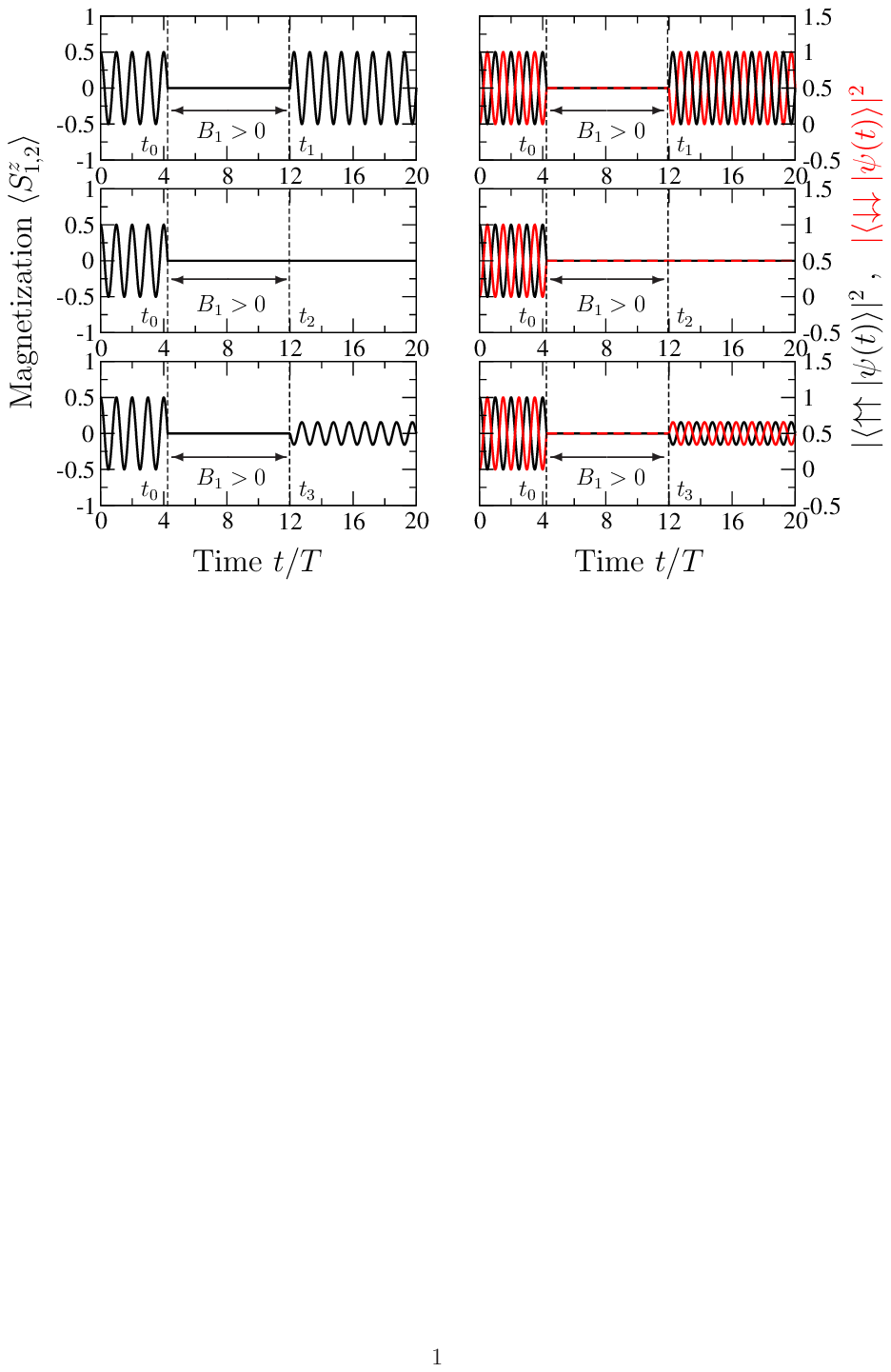}
  \end{center}
  \caption{(color online) Second scenario: The left column shows the
    magnetization $\langle S_z \rangle$, and the right column the
    probabilities of finding the system in one of the classical spin
    spiral states. The magnetization, as well as the probabilities, are
    given in local coordinates. In all three cases, $|\psi(t)\rangle$ starts
    in the oscillatory state $|\psi_{+i} \rangle =
    (|\Psi_{S,\uparrow}\rangle + i |\Psi_{S,\downarrow}\rangle)/\sqrt{2}$. The
    oscillation stops by switching on a magnetic field $B_1$ at a time
    $t_0$ when the system is in a symmetric superposition of both
    classical spiral states. The magnetic field $B_1$ is switched off
    at times $t_1$, $t_2$, respectively $t_3$. Please note that $t_1$,
    $t_2$, and $t_3$ are different times. The horizontal arrow shows the time when the local magnetic field is switched on $B_1 >
    0$. Outside this time window, the magnetic field is switched
    off. The upper row corresponds to the situation when $B_1$ is
    switched off at time $t_1 = (n\cdot \pi/2)/\omega_B$, when
    $|\psi(t)\rangle$ is one of the states (\ref{PSIBII}), this leads
    to the return to the oscillation between the two classical spiral
    states. In the middle row, the magnetic field $B_1$ is switched
    off at time $t_2 = ([2n+1]\cdot \pi/4 )/ \omega_B$, when
    $|\psi(t)\rangle$ is in one of the states (\ref{PSIBII}). In these cases, the dimer is either in the ground state
    $|\mathrm{GS}\rangle$ or in the excited state $|\mathrm{E}\rangle$ after the field is switched off. The lower row shows a situation where $B_1$ is switched off at time $t_3 \neq t_{1,2}$ when 
    $|\psi(t)\rangle$ is between the symmetrical states (\ref{PSIB})
    and (\ref{PSIBII}). The time $t$ is taken in units of the single
    oscillation time $T = \pi \hbar / J_x$ between the two classical
    spiral states if the magnetic field $B_1$ is switched off.}   
  \label{f:pic4} 
\end{figure*}
\begin{figure}[ht]
  \begin{center}
    \includegraphics*[width=3.75cm,bb = 200 0 500 800,angle=90]{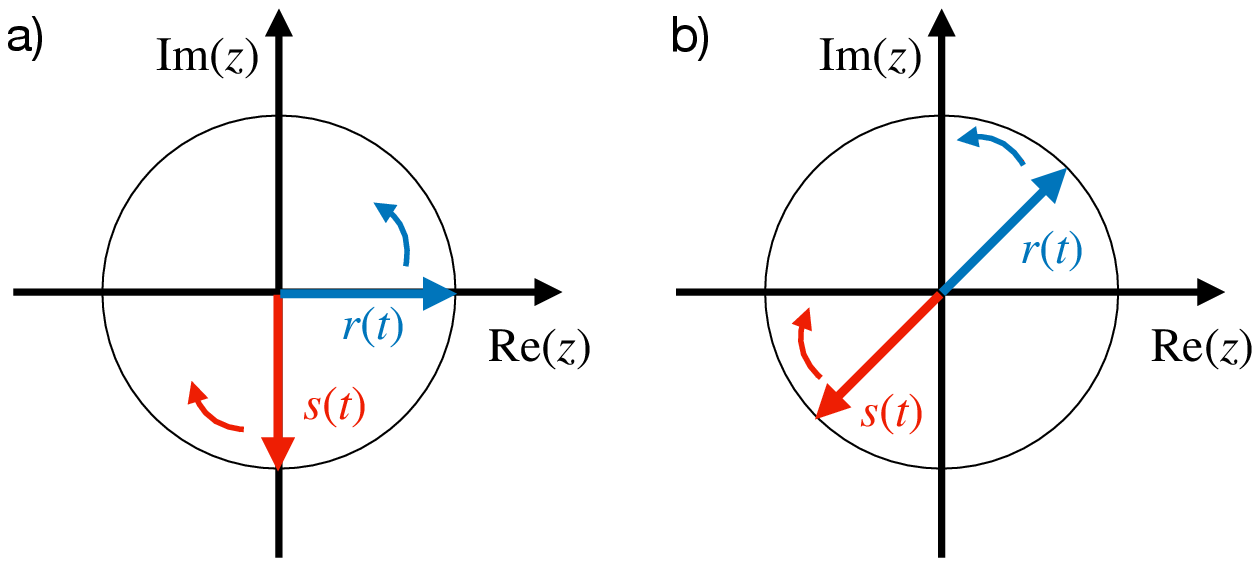}
  \end{center}
  \caption{(color online) Countercyclically rotation of the coefficients
    $r(t) = \exp(iB_1t/\hbar)$ and $s(t) =
    -i\exp(-iB_1t/\hbar)$ of $|\psi(t)\rangle$, given by
    Eq.~(\ref{B1Os}), at different times a) $t = 0$, respectively, 
    b) $t = \frac{\pi\hbar}{4B_1}$.}  
  \label{f:pic5} 
\end{figure}

The second interesting scenario is when we stop the oscillation when the dimer is in one of the symmetrical superposition states:
\begin{eqnarray} \label{PSIpmi}
  |\psi_{\pm i}\rangle = \frac{1}{\sqrt{2}}\Big(|\Psi_{S,\uparrow}\rangle \pm i
  |\Psi_{S,\downarrow}\rangle  \Big) \;.
\end{eqnarray}
These states are characterized by the fact that there is no magnetic
order $\langle \vec{S}_1 \rangle = \langle \vec{S}_2 \rangle = 0$, and the two spins are maximally entangled.
In local coordinates, these quantum states are written as:
\begin{eqnarray} \label{StateStartB1}
  |\psi_{\pm i}\rangle =
  \frac{1}{\sqrt{2}}\Big(|\!\uparrow\uparrow\rangle \pm i 
  |\!\downarrow\downarrow\rangle  \Big) \;.
\end{eqnarray}
The energy of these states is $E = -\tilde{J}$. This is 
energetically in the middle of the energy gap between the ground state
$|\mathrm{GS}\rangle$ and the excited state
$|\mathrm{E}\rangle$. Fig.~\ref{f:pic4} shows the possible outcomes
of such experiments.

The interesting aspect of this scenario is that the quantum state
changes under the influence of the external magnetic field $B_1$. This
is not the case for the classical spiral states
$|\Psi_{S,\uparrow}\rangle$ and $|\Psi_{S,\downarrow}\rangle$. If we assume that the magnetic field is dominating $B_1 \gg J_x$ and
the initial state is $|\psi_{- i}\rangle$, the dynamics of the quantum state, in local coordinates, is given by:  
\begin{eqnarray}  \label{B1Os}
|\psi(t)\rangle = \frac{1}{\sqrt{2}}\left(
e^{\frac{iB_1t}{\hbar}}|\!\uparrow\uparrow\rangle -i
e^{\frac{-iB_1t}{\hbar}}|\!\downarrow\downarrow\rangle\right)e^{\frac{i
\tilde{J}t}{\hbar}} \;.
\end{eqnarray}
What is relevant here is the expression in the parenthesis. The coefficients of both spiral states are described by complex
exponential functions, which describe rotations in the complex plane
with frequency $\omega_B = B_1/\hbar$. Here, both coefficients rotate
countercyclically, see Fig.~\ref{f:pic5}. If the phase in Eq.~({\ref{B1Os})
$\omega_B t$ is of the form $\omega_B t = (2n+1)\cdot
\pi/4$, with $n \in \mathbb{Z}$, the quantum state is in one of the following states:  
\begin{subequations} \label{PSIB}
  \begin{eqnarray} 
    |\psi_{\alpha} \rangle &=& \frac{\alpha}{\sqrt{2}}
    \Big(|\!\uparrow\uparrow\rangle +
    |\!\downarrow\downarrow\rangle \Big)  \;, \label{PSIB12} \\
    |\psi_{\beta} \rangle &=& \frac{\beta}{\sqrt{2}}  \Big(|\!\uparrow\uparrow\rangle -
    |\!\downarrow\downarrow\rangle \Big) \;. \label{PSIB34}  
  \end{eqnarray}
\end{subequations}
As long as the magnetic field $B_1$ is switched on, these states have energy $E = -\tilde{J}$. As soon as the magnetic field $B_1$ is
switched off, $|\psi_{\alpha} \rangle$ becomes the ground state
$|\mathrm{GS}\rangle$, and $|\psi_{\beta} \rangle$ become the excited state $|\mathrm{E}\rangle$. In both cases, there are overall
phases $\alpha = \sqrt{2}e^{\frac{i(4n+3)\pi}{4}}$, respectively, 
$\beta = \sqrt{2}e^{\frac{i(4n+1)\pi}{4}}$, which have no further meaning. If the phase $\omega_Bt$ is of the form
$\omega_B t = n\cdot \pi/2$, $n \in \mathbb{Z}$, the quantum state
$|\psi(t)\rangle$ is in one 
of the following states:  
\begin{subequations} \label{PSIBII}
  \begin{eqnarray} 
    |\psi_{\gamma} \rangle &=& \frac{\gamma}{\sqrt{2}}\Big(
    |\!\uparrow\uparrow\rangle + i
    |\!\downarrow\downarrow\rangle \Big) \;, \label{PSIB56} \\
    |\psi_{\delta} \rangle &=& \frac{\delta}{\sqrt{2}}\Big( 
    |\!\uparrow\uparrow\rangle - i
    |\!\downarrow\downarrow\rangle \Big)\;. \label{PSIB78}  
  \end{eqnarray}
\end{subequations}
These states are up to irrelevant overall phases,$\gamma =
e^{\frac{i(2n+1)\pi}{2}}$, respectively, $\delta =
e^{\frac{i(2n)\pi}{2}}$ , equal to the quantum states given by 
Eq.~(\ref{StateStartB1}). 

As soon as the external magnetic field is switched off, the dimer will
start oscillating again. However, this oscillation strongly depends on
which quantum state the system is in at the moment the magnetic field
$B_1$ is switched off. If the dimer is in one of the states given by
Eq.~(\ref{PSIBII}), the oscillation between the two classical spiral
states, $|\Psi_{S,\uparrow}\rangle$ and $|\Psi_{S,\downarrow}\rangle$,
will be recovered. Which of the two classical spiral states is reached
first after the reinitialization of the oscillation depends on the
quantum state of the system at the time when the magnetic field $B_1$ is
switched off. If the quantum state is $|\psi_\gamma\rangle$, then $|\Psi_{S,\uparrow}\rangle$ will be reached first, but if the
quantum state is $|\psi_\delta\rangle$, then
$|\Psi_{S,\downarrow}\rangle$ will be reached first. In other words,  
the quantum states $|\psi_{\gamma}\rangle$ and $|\psi_{\delta}\rangle$ lead
to opposite oscillations. This is independent of which
classical spiral state the dimer was in before the magnetic field
was turned on. In this scenario, the oscillation continues as if it
had not been stopped. On the other hand, if the dimer is in one of the
states given in Eq.~(\ref{PSIB}), the dimer will be either in the ground
state $|\mathrm{GS}\rangle$ or the excited state $|\mathrm{E}\rangle$ after the magnetic field $B_1$ is switched off. No oscillation will
occur in these two cases, and the dimer stays entangled. When
switching on the magnetic field $B_1$ again, the oscillation described
by Eq.~(\ref{B1Os}) will be reinitialized, and it is possible to recover the
oscillation between the two classical spiral states after the
magnetic field is switched off. 

So far, we have discussed the scenarios where we can reproduce the
ground state $|\mathrm{GS}\rangle$, the excited state
$|\mathrm{E}\rangle$, and the superposition states $|\psi_{\pm
  i}\rangle$. These states have energies $E = -\tilde{J} \pm J_x$ 
and $E = -\tilde{J}$. If the quantum state $|\psi(t)\rangle$, at the
time when the magnetic field $B_1$ is switched off, is not one of the
discussed quantum states, Eqns.~(\ref{PSIB}) or (\ref{PSIBII}), its
energy $E$ will be between $E_{min} = -\tilde{J} - J_x$ and $E_{max} =
-\tilde{J} + J_x$. In these cases, one finds an oscillation with a reduced
amplitude; this is shown in the lower row of Fig.~\ref{f:pic4}.     
 
It is important to emphasize that the timing of switching off the magnetic field $B_1$ is crucial. The quantum state of the dimer at the moment $B_1$ is turned off determines whether the system remains in a fully entangled state, resumes coherent oscillation between the two classical spiral states, or enters an oscillation with reduced amplitude. This controllability highlights the potential of spin-spiral dimers for applications in quantum computing and quantum information processing.

The external magnetic field $B_1$, which needs to be local, can be created, e.g., by a spin-polarized magnetic tip or by the spin current
of such a magnetic tip. Alternatively,  experiments with ultra-cold atoms and lasers are also possible. 
\begin{figure}[ht]
  \begin{center}
    \includegraphics*[width=7cm,bb = 75 460 540 765]{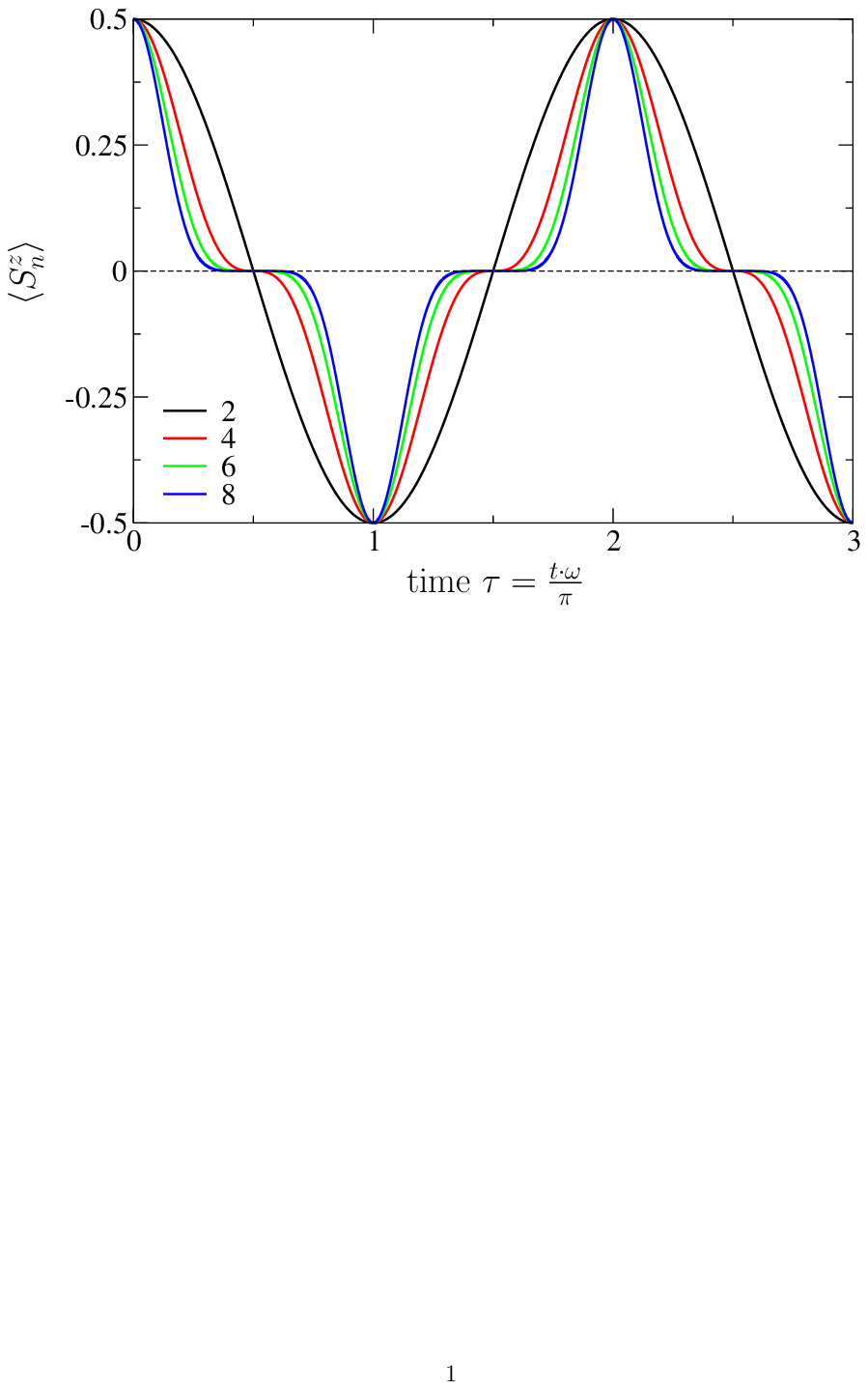}
  \end{center}
  \caption{(color online) Spin expectation values $\langle S_n^z \rangle$ as a function of time $\tau$ for spin-spiral chains with $N = 2$, $4$, $6$, and $8$ spins. The spin expectation values are in local coordinates. In global coordinates, $\langle S_n^z \rangle$ corresponds to the amplitude of $\langle \vec{S}_n \rangle$.}   
  \label{f:pic6} 
\end{figure}
\begin{figure*}[ht]
  \begin{center}
    \includegraphics*[width=7cm,bb = 75 450 540 765]{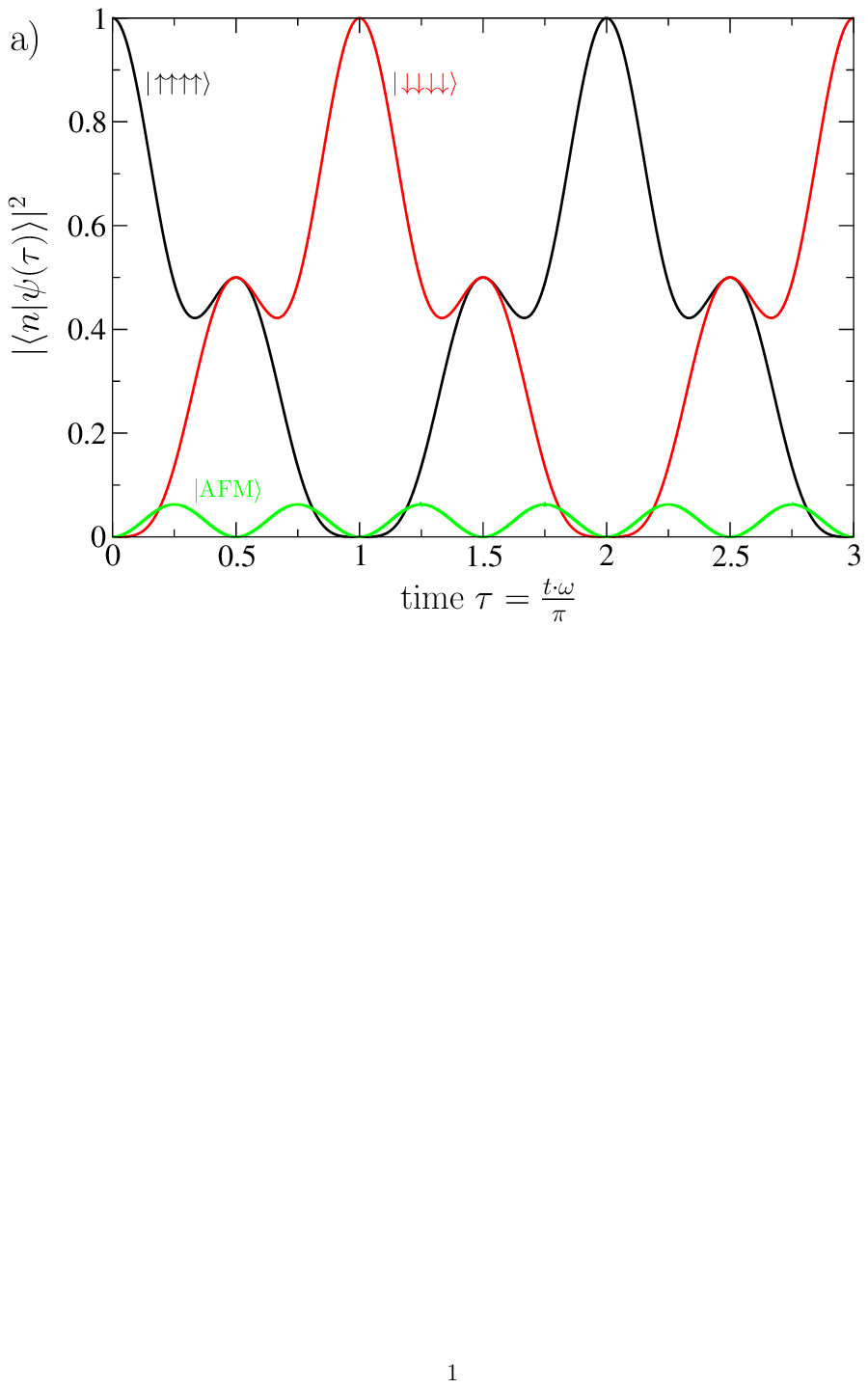}
    \includegraphics*[width=7cm,bb = 75 450 540 765]{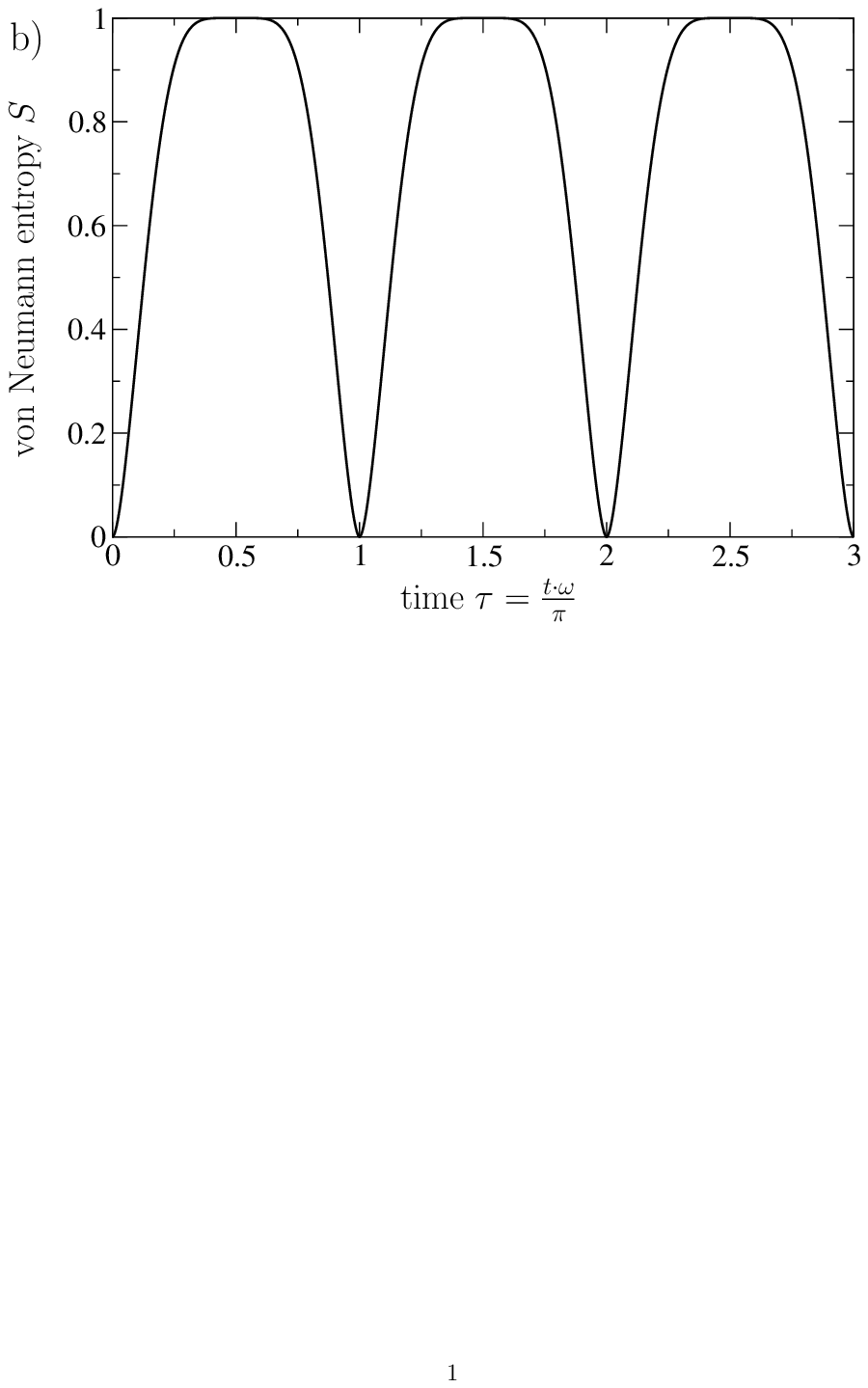}
  \end{center}
  \caption{(color online) Probability to find the basis state $|n\rangle$ and the von Neumann entropy $S$ as a function of time $\tau$. The description is in local coordinates. The ferromagnetic basis states $|\!\uparrow\uparrow\uparrow\uparrow \rangle$ and $|\!\downarrow\downarrow\downarrow\downarrow \rangle$ represent the classical spin spirals in global coordinates. The $|\mathrm{AFM}\rangle$ are all possible basis states corresponding to antiferromagnetic spin spirals.}   
  \label{f:pic7} 
\end{figure*}

So far, we have discussed the dynamics of a spin spiral dimer and the possibility of manipulating the quantum state by stopping the oscillation. The question that now arises is whether this dynamic can also be observed with longer chains with a spin spiral. In the previous part of the paper, the dynamics have been studied using analytical calculations. From now on, the dynamics will be described by numerically solving the time-dependent Schr\"odinger equation. It turns out that a similar oscillation, as described in the case of the dimer, can only be seen if the spin chain has an even number of spins. All examined spin chains with an even number of spins show no magnetic order $\langle \vec{S}_n \rangle = 0$. In contrast, spinets with an odd number of spins show a magnetic spiral order, only with reduced spin expectation values $|\langle \vec{S}_n \rangle| < 1/2$. This is analogous to the observations with antiferromagnetic chains \cite{machensPRB13}. As mentioned, an oscillation between the two inverse classical spin-spiral states can only be observed for spin spirals with an even number of spins. For the spin dimer, we always find an oscillation between the two classical spin-spiral states. For $N \geq 4$ spins, such an oscillation only occurs if $D \ll J$. In local coordinates, this constraint means $J_x \ll {\tilde J}$. In global coordinates, $D \ll J$ means a tiny spiral angle $\theta$, such as one degree. The dynamics become disordered if the condition $D \ll J$ is not fulfilled. Fig. \ref{f:pic6} shows the spin expectation values $\langle S_n^z \rangle$ in local coordinates for chains with 2, 4, 6, and 8 spins. The dynamics of the spin expectation values in global coordinates can be achieved by a rotation. Within the simulations, the ratio of ${\tilde J}/J_x = 1/10000$ was assumed. In global coordinates, this corresponds to a $D/J$ of $\sqrt(2)/100 = 0.01414$ and a spiral angle of less than 1 degree. The numerical results obtained for the two-spin chain (dimer) are consistent with the analytical results derived previously. The initial states of the simulations are the ferromagnetic states with N spins up $| \!\uparrow \ldots \uparrow \rangle$, these are states in local coordinates and correspond to the classical ferromagnetic spin spirals. These initial states are energetically equal to the inverse spiral states. These spiral states correspond to the ferromagnetic states in local coordinates with N spins down $| \!\downarrow \ldots\downarrow \rangle$. The dynamics is then an oscillation between the initial and inverse spiral states. The corresponding spin expectation values are shown in Fig. \ref{f:pic6}. When the number of spins, $N$, increases, some significant changes in the magnetization curves can be seen. On the one hand, plateaus occur in the middle of the oscillation at $\langle S_n^z \rangle = 0$. This plateau does not exist for the case of $N = 2$ (dimer) and becomes more pronounced as the number $N$ of spins increases. On the other hand, the oscillation frequency $\omega$ decreases with an increasing number of spins. The frequency is given as $\omega = \frac{4J_x}{N\hbar}$. 

The 4-spin chain shall be described in more detail to get a better understanding of the oscillation dynamics. The probabilities of finding the basis states $|n\rangle$ describing the quantum state $|\psi(t)\rangle$, and the von Neumann entropy $S$ are shown in Fig. \ref{f:pic7}. The description is in local coordinates.

In the scenario of a 4-spin chain, the spin expectation values $\langle S_n^z \rangle$ are oscillating between the two classical spiral states $|\!\uparrow\uparrow\uparrow\uparrow \rangle$ and $|\!\downarrow\downarrow\downarrow\downarrow \rangle$. This results in an oscillation between $\langle S_n^z \rangle = 1/2$ and $\langle S_n^z \rangle = -1/2$. In global coordinates, $\langle S_n^z \rangle$ can be seen as the amplitudes of the spin expectation values $\langle {\cal S}_n^x \rangle$ and $\langle {\cal S}_n^z \rangle$. As mentioned above, a plateau around $\langle S_n^z \rangle = 0$ is formed during this oscillation. If we now examine the temporal evolution of the quantum state, particularly the probabilities of the base states' occurrence (see Fig. \ref{f:pic7}a)), the oscillation between the two classical spiral states becomes evident. These states are represented in local coordinates by the ferromagnetic configurations $|\!\uparrow\uparrow\uparrow\uparrow\rangle$ and $|\!\downarrow\downarrow\downarrow\downarrow\rangle$. In addition to these two base states, the antiferromagnetic base states $|\mathrm{AFM}\rangle = |\!\uparrow\downarrow\uparrow\downarrow\rangle$, $|\!\downarrow\uparrow\downarrow\uparrow\rangle$, $|\!\uparrow\uparrow\downarrow\downarrow\rangle$, $|\!\uparrow\downarrow\downarrow\uparrow\rangle$, $|\!\downarrow\downarrow\uparrow\uparrow\rangle$, and $|\!\downarrow\uparrow\uparrow\downarrow\rangle$ also occur. These basis states have to be seen as the classical antiferromagnetic spin-spiral states. The global states occur by rotating the spins by half the spiral angle. The remaining eight basis states are irrelevant in this context. Special attention should be given to the quantum states at $\tau = 1/2$ and $\tau = 3/2$. At $\tau = 1/2$, the quantum state is $|\psi(t)\rangle = \frac{1}{\sqrt{2}}[|\!\uparrow\uparrow\uparrow\uparrow\rangle - i |\!\downarrow\downarrow\downarrow\downarrow\rangle]$. The simulation reveals that at $\tau = 3/2$, the quantum state evolves to $|\psi(t)\rangle = \frac{1}{\sqrt{2}}[|\!\uparrow\uparrow\uparrow\uparrow\rangle + i |\!\downarrow\downarrow\downarrow\downarrow\rangle]$. This result aligns with the analytical calculation of the $N = 2$ spin chain and can also be found for the $N = 6$ and $N = 8$ spin spiral. Furthermore, the von Neumann entropy $S$, which quantifies the entanglement of the quantum state $|\psi(t)\rangle$, exhibits a periodic alternation between $S=0$ (unentangled classical spiral states) and $S=1$ (maximally entangled symmetrical superpositions of the classical spiral states), as shown in figure \ref{f:pic7}b). This alternation between non-entangled and entirely entangled was also observed in the $N = 2$ spin spiral (dimer). The key difference lies in the duration of total entanglement, which corresponds to the plateau in $\langle S_n^z \rangle$. This observation suggests that the analytically described method for manipulating the quantum state can also be extended to longer chains. Notably, the plateau offers an advantage by lengthening the period of maximum entanglement, providing more time for potential manipulation of the quantum state.

In summary, this article investigates the spin dynamics of the ground state configuration of a spin spiral dimer under the influence of a magnetic field $\vec{B}_1$ that acts on a single spin. By adiabatically varying $\vec{B}_1$, the non-magnetic ground state can transition to a classical spiral state, which then oscillates periodically to its inverted configuration. This oscillation can be halted at any point, leading to two intriguing scenarios:
\begin{enumerate}
    \item Stopping the oscillation when $|\psi(t)\rangle$ is in one of the classical spiral states allows precise control of the oscillation.
    \item Stopping the oscillation when $|\psi(t)\rangle$ is in a symmetric superposition of the two classical spiral states induces a periodic change in the coefficients of the spiral states due to the external magnetic field. Here, the timing of switching off the magnetic field is critical, determining whether the system resumes oscillation between the classical spiral states $|\Psi_{S,\uparrow}\rangle$ and $|\Psi_{S,\downarrow}\rangle$, or settles into one of the maximally entangled eigenstates $|\mathrm{GS}\rangle$ and $|\mathrm{E}\rangle$. These two scenarios hold potential for applications in quantum computing and information processing.
\end{enumerate}
 
While the study focuses on ferromagnetic interactions, similar dynamics occur for antiferromagnetic spin dimers. Furthermore, this concept can be extended to longer spin spiral chains, provided the system has an even number of spins and a small Dzyaloshinsky-Moriya interaction $D \ll J$. In such cases, the spin chains exhibit oscillations analogous to those of the dimer. The primary distinction is the emergence of a plateau in the spin expectation values $\langle S_n^z \rangle$, corresponding to a state of complete entanglement. This plateau, which extends the duration of maximal entanglement, grows with the number of spins, providing an extended window for manipulating the quantum state in a manner similar to the dimer.

\begin{acknowledgments}
R. Wieser acknowledges the financial support provided by
the Startup Foundation for Introducing Talent of NUIST (2018r043).  
\end{acknowledgments}

\bibliography{Cite30.11}

\end{document}